\documentclass[12pt]{article}
\usepackage{amssymb}
\usepackage{amsmath}
\usepackage{apacite}
\usepackage{latexsym}
\usepackage{epsfig}
\usepackage{graphicx}
\usepackage{subfigure}
\usepackage{wasysym}
\usepackage{threeparttable}
\usepackage{natbib}
\usepackage{color}
\usepackage{epstopdf}
\usepackage{bm}
\usepackage{float}
\usepackage{todonotes}
\usepackage{verbatim}

\usepackage{hyperref}

\def\boxit#1{\vbox{\hrule\hbox{\vrule\kern6pt
          \vbox{\kern6pt#1\kern6pt}\kern6pt\vrule}\hrule}}

\def\bse{\begin{eqnarray*}}
\def\ese{\end{eqnarray*}}
\def\be{\begin{eqnarray}}
\def\ee{\end{eqnarray}}
\def\bq{\begin{equation}}
\def\eq{\end{equation}}
\def\bse{\begin{eqnarray*}}
\def\ese{\end{eqnarray*}}

\begin{document}

\thispagestyle{empty} 
\baselineskip=28pt

\begin{center}
{\LARGE{\bf Conjugate Bayesian Unit-level Modeling of Count Data Under Informative Sampling Designs}}

\end{center}

\baselineskip=12pt

\vskip 2mm
\begin{center}
Paul A. Parker\footnote{(\baselineskip=10pt to whom correspondence should be addressed)
Department of Statistics, University of Missouri,
146 Middlebush Hall, Columbia, MO 65211-6100, paulparker@mail.missouri.edu},

   Scott H. Holan\footnote{\baselineskip=10pt Department of Statistics, University of Missouri,
146 Middlebush Hall, Columbia, MO 65211-6100, holans@missouri.edu}\,\footnote{\baselineskip=10pt U.S. Census Bureau, 4600 Silver Hill Road, Washington, D.C. 20233-9100, scott.holan@census.gov},

 and Ryan Janicki\footnote{\baselineskip=10pt U.S. Census Bureau, 4600 Silver Hill Road, Washington, D.C. 20233-9100}
\\
\end{center}

\vskip 4mm
\baselineskip=12pt 
\begin{center}
{\bf Abstract}
\end{center}

Unit-level models for survey data offer many advantages over their area-level counterparts, such as potential for more precise estimates and a natural benchmarking property. However two main challenges occur in this context: accounting for an informative survey design and handling non-Gaussian data types. The pseudo-likelihood approach is one solution to the former, and conjugate multivariate distribution theory offers a solution to the latter. By combining these approaches, we attain a unit-level model for count data that accounts for informative sampling designs and includes fully Bayesian model uncertainty propagation. Importantly, conjugate full conditional distributions hold under the pseudo-likelihood, yielding an extremely computationally efficient approach. Our method is illustrated via an empirical simulation study using count data from the American Community Survey public-use microdata sample.

\baselineskip=12pt
\par\vfill\noindent
{\bf Keywords:} Bayesian, Conjugate multivariate distribution, Count data, Pseudo-likelihood, Public-use microdata, Small area estimation, Survey data.
\par\medskip\noindent
\clearpage\pagebreak\newpage \pagenumbering{arabic}
\baselineskip=24pt


\section{Introduction}

Statistical estimates from survey samples have traditionally been obtained via design-based estimators \citep{lohr09}. In many cases, these estimators tend to work well for quantities such as population totals or means, but can fall short as sample sizes become small. In today's ``information age," there is a strong demand for more granular estimates. The Small Area Income and Poverty Estimates program (SAIPE) and the Small Area Health Insurance Estimates program (SAHIE) are two examples that rely on American Community Survey (ACS) data, where granularity is essential \citep{luery11, bau18}. Both of these programs require annual estimates at the county level for the entire United States. Many counties exhibit extremely small sample sizes, or even a sample size of zero. In these cases, design-based estimation is inadequate and model-based estimation becomes necessary.

Models for survey data can be either at the area level or the unit level. Area-level models typically use design-based estimators as the response and tend to smooth the estimates in some fashion. These models often use area-level random effects to induce smoothing, and thus a common application is small area estimation (see, for example, \citet{porter15} and the references therein). \citet{rao15} provide a recent overview of many of the current area-level models that are available. One issue with area-level models is that estimates at a finer geographic scale may not aggregate to estimates at coarser spatial resolutions, thereby producing inconsistencies.

Unit-level models include individual response values from the survey units as response variables rather than the area-level design-based estimators. The basic unit-level model was introduced by \citet{bat88} in order to estimate small area means. One advantage of unit-level modeling is that the response value can be predicted for all units not contained in the sample, and thus estimates for finite population quantities aggregate naturally. In addition, unit-level models have the potential to yield more precise estimates than area-level models \citep{hid16}. When modeling survey data at the unit level, the response is often dependent on the sample selection probabilities. This scenario is termed {\it informative sampling}, and it is critical to incorporate the design information into the model in order to avoid biased estimates \citep{pfe07}. Various approaches exist for incorporating an informative design into a model formulation. \citet{little12} suggests the use of design variables in the model. For simple survey designs, this may work well, but can become infeasible for complex survey designs. \citet{si15} and \citet{van16} both use nonparametric regression techniques on the survey weights. These types of techniques do not require any knowledge of the survey design, though they can be difficult to implement in the presence of covariates. Finally, the often used pseudo-likelihood approach \citep{ski89, bin83} exponentially weights each unit's likelihood contribution by the corresponding survey weight. In this way, the sample data model is adjusted to better match the population distribution. See \citet{par19} for a recent review. 

In addition to the issues that arise due to informative sampling, many variables found within survey data are non-Gaussian in nature, which may induce modeling difficulties. Two examples present in American Community Survey (ACS) data are a binary indicator of health insurance coverage and a count of the number of bedrooms within a household. The SAE setting is often aided by the use of area-level and/or unit-level random effects, which is commonly done using Bayesian hierarchical modeling with a latent Gaussian process (LGP) \citep{cress11,gelf16}. In the presence of non-Gaussian data, LGP models lead to non-conjugate full conditional distributions that can be difficult to sample from. \citet{brad17} provide a solution to this problem by appealing to a class of multivariate distributions that are conjugate with members of the natural exponential family.

We introduce a modeling framework for dealing with unit-level count data under informative sampling by using Bayesian hierarchical modeling to account for complex dependence structures (e.g., in space and time), and relying on the distribution theory provided by \cite{brad17} for computationally efficient sampling of the posterior distribution. To account for informative sampling, we use a Bayesian pseudo-likelihood \citep{sav16}. In Section~\ref{sec:meth} we introduce and discuss our modeling approach. Section~\ref{sec:sim} considers a simulation study comparing our methodology to that of two competing estimators. Finally, we provide discussion in Section~\ref{sec:disc}.

\section{Methodology}\label{sec:meth}
\subsection{Informative Sampling}
\citet{par19} review current approaches to unit-level modeling under informative sampling. Some of the general approaches include incorporating the design variables into the model \citep{little12}, regression on the survey weights \citep{si15, van16}, and joint modeling of the response and weights \citep{pfe07,novelo17}. Another general approach is to use a weighted pseudo-likelihood.

Let \(\mathcal{U} = \{1, \dots, N\}\) be an enumeration of the units in the
population of interest, and let \(\mathcal{S} \subset \mathcal{U}\) be the observed,
sampled units, selected with probabilities \(\pi_i = P(i \in \mathcal{S})\).  Let
\(y_i\) be a variable of interest associated with unit \(i \in \mathcal{U}\).  Our
goal is inference on the finite population mean \(\bar{y} = \sum_{i \in \mathcal{U}}
y_i / n\).  Suppose a model, \(f(y_i \mid \theta)\), conditional on a vector of
unknown parameters, \(\theta\), holds for the units \(y_i\) for \(i \in
\mathcal{U}\).  If the survey design is informative, so that the selection
probabilities, \(\pi_i\), are correlated with the response variables, \(y_i\), the
model for the nonsampled units will be different from the model for the sampled
units, making inference for the finite population mean challenging. Often, the reported survey weights, $w_i=1/ \pi_i,$ are used to account for the survey design.

The pseudo-likelihood (PL) approach, introduced by \citet{ski89} and \citet{bin83}, uses the survey weights to re-weight the likelihood contribution of sampled units. The pseudo-likelihood is given by
\begin{equation}\label{E: pseudoLogLikelihood}
  \prod_{i \in \mathcal{S}}  f( y_i \mid \boldsymbol{\theta})^{w_i},
\end{equation} 
where $y_i$ is the response value for unit $i$ in the sample $\mathcal{S}$. In (\ref{E: pseudoLogLikelihood}), the  vector of model parameters is denoted by $\boldsymbol{\theta}$ and the survey weight for unit $i$ is denoted by $w_i$. For frequentist estimation, the PL can be maximized via maximum likelihood techniques, whereas \citet{sav16}  use the PL in a Bayesian setting for general models. Modeling under a Bayesian pseudo-likelihood induces a pseudo-posterior distribution
\begin{eqnarray*}
\hat{\pi}(\bm{\theta} | \mathbf{y}, \mathbf{\tilde{w}}) \propto \left\{ \prod_{i \in \mathcal{S}} f(y_{i} | \bm{\theta})^{\tilde{w}_{i}} \right\} \pi (\bm{\theta}),
\end{eqnarray*}
where $\tilde{\mathbf{w}}$ represents the weights after being scaled to sum to the sample size. This scaling is done in order to keep the asymptotic amount of information the same as the regular likelihood case, and prevent under-estimation of the standard errors, since the weights act as frequency weights \citep{sav16}. It was shown by \citet{sav16}, that the
pseudo-posterior distribution converges to the population posterior distribution,
justifying the use of the pseudo-posterior distribution for inference on the
nonsampled units.

The PL approach is geared towards parameter estimates and not necessarily estimates of finite population quantities. Nevertheless, in this setting (and others), poststratification is a general technique that can be used to create finite population quantity estimates. The general idea is to use a model to predict the response value for all unsampled units in the population, effectively generating a population that can be used to extract any desired estimates. \citet{little93} gives an overview of poststratification, whereas \citet{gelman97} and \citet{park06} develop the idea of poststratification under Bayesian hierarchical models.

\subsection{Modeling Non-Gaussian Data}
Many of the variables collected from complex surveys are non-Gaussian. For example, binary indicators and count data are both very common in survey data, but cannot be modeled at the unit level under a Gaussian response framework. As such, this can lead to computational issues when dependence structures are introduced.

Bayesian hierarchical modeling is commonly used to model complex dependence structures such as those found in sample surveys. These models often consist of a data stage that models the response, a process stage, and a prior distribution over model parameters. Traditionally, a latent Gaussian process is used to model the process stage; for example see \citet{cress11}. \citet{gelf16} review the use of Gaussian process modeling in spatial statistics, and \citet{brad15} develop a general LGP framework that handles multivariate responses as well as complex spatio-temporal dependence structures.

In a Bayesian setting, when the response variable is also Gaussian, Gibbs sampling can be implemented to efficiently sample from the posterior distribution. Unfortunately, when dealing with non-Gaussian data, a Metropolis-Hastings type step may be necessary within the Markov chain Monte Carlo algorithm. Consequently, this algorithm must be tuned and can lead to poor mixing, especially in high-dimensions. Because many survey variables are inherently non-Gaussian, the Gaussian process framework is not ideal in many survey data scenarios.

\citet{brad17, brad18, brad18b} incorporate new distribution theory to create a set of Bayesian hierarchical models that maintain conjugacy for any response variable contained in the natural exponential family. This includes Poisson, Bernoulli, Binomial, and Gamma random variables, among others, and thus offers a very general modeling framework that maintains computational efficiency.

In this work we consider the distribution theory for Poisson responses specifically, in order to model count survey data. \citet{brad17} further consider the Negative Binomial case, but state that modeling can be more challenging in this scenario. They suggest that Negative Binomial data may be alternatively modeled as Poisson, and inclusion of random effects can help to model overdispersion. 

Each natural exponential family response type is shown to be conjugate with a class of distributions referred to as the conjugate multivariate (CM) distribution. For a Poisson response, the CM distribution is the multivariate log-Gamma (MLG) distribution with probability density function (PDF)
\begin{equation}
    \hbox{det}(\bm{V}^{-1})
    \left\{ \prod_{i=1}^n \frac{\kappa_i^{\alpha_i}}{\Gamma(\alpha_i)}\right\}
    \hbox{exp}\left[\bm{\alpha}' \bm{V}^{-1}(\bm{Y- \mu}) -
    \bm{\kappa}' \hbox{exp}\left\{\bm{V}^{-1}(\bm{Y- \mu}) \right\}\right],
\end{equation} denoted by $\hbox{MLG}(\bm{\mu}, \mathbf{V}, \bm{\alpha}, \bm{\kappa})$. The MLG distribution is easy to simulate from using the following steps:
\begin{enumerate}
        \item Generate a vector $\mathbf{g}$ as $n$ independent Gamma random variables with shape $\alpha_i$ and rate $\kappa_i$, for $i=1,\ldots,n$
        \item Let $\mathbf{g}^*=\hbox{log}(\mathbf{g})$
        \item Let $\mathbf{Y}=\mathbf{V g}^* + \bm{\mu}$
        \item Then $\mathbf{Y} \sim \hbox{MLG}(\bm{\mu}, \mathbf{V}, \bm{\alpha}, \bm{\kappa}).$
    \end{enumerate} 

Bayesian inference with Poisson data and MLG prior distribution also requires simulation from the conditional multivariate log-Gamma distribution ($\hbox{cMLG}$). Letting $\mathbf{Y} \sim \hbox{MLG}(\bm{\mu},  \mathbf{V}, \bm{\alpha}, \bm{\kappa})$, \citet{brad18} show that $\mathbf{Y}$ can be partitioned into $(\mathbf{Y_1}', \mathbf{Y_2}')'$, where $\mathbf{Y_1}$ is $r$-dimensional and $\mathbf{Y_2}$ is $(n-r)$-dimensional. The matrix $\mathbf{V}^{-1}$ is also partitioned into $\left[\mathbf{H \; B}  \right]$, where $\mathbf{H}$ is an $n \times r$ matrix and $\mathbf{B}$ is an $ n \times (n - r)$ matrix. Then 
$$\bm{Y_1} | \bm{Y_2} = \bm{d}, \bm{\mu}^*, \bm{H}, \bm{\alpha}, \bm{\kappa} \sim \hbox{cMLG}(\bm{\mu}^*, \bm{H}, \bm{\alpha}, \bm{\kappa}; \Psi)$$ with density
\begin{equation}
    M \hbox{exp} \left\{\bm{\alpha}' \bm{H Y_1} 
     - \bm{\kappa}' \hbox{exp}(\bm{H Y_1} - \bm{\mu}^*)\right\} I\left\{(\bm{Y_1}' , \bm{d}')' \in \mathcal{M}^n \right\},
\end{equation} where $\bm{\mu}^*=\mathbf{V}^{-1}\bm{\mu} - \mathbf{Bd}$, and $\mathit{M}$ is a normalizing constant. It is also easy to sample from the cMLG distribution when doing Bayesian analysis by using a collapsed Gibbs sampler \citep{liu94}. \citet{brad17} show that this can be done by drawing $(\mathbf{H}'\mathbf{H})^{-1}\mathbf{H}'\mathbf{Y}$, where $\mathbf{Y}$ is sampled from $\hbox{MLG}(\bm{\mu}, \mathbf{I}, \bm{\alpha}, \bm{\kappa})$.

\subsection{Pseudo-likelihood Poisson Multivariate log-Gamma Model}
In order to use the conjugate multivariate distribution theory of \citet{brad17} in a survey setting for count data under informative sampling, we replace the Poisson likelihood with a survey weighted pseudo-likelihood. Under the unweighted setting, the likelihood contribution to the posterior is proportional to 
\begin{equation*}
    \prod_{i \in \mathcal{S}} \hbox{exp}\left\{Z_i Y_i -  b_i \hbox{exp}(Y_i)  \right\} = 
    \hbox{exp}\left\{\bm{Z' Y} - \mathbf{b}' \hbox{exp}(\bm{Y})  \right\},
\end{equation*}
 with $\mathbf{Z}$ representing a vector of response variables, and $\mathbf{Y}$ representing a parameter vector, which will later be modeled using the MLG distribution. The parameter $b_i=1$ for the Poisson case. This expression is proportional to the product of Poisson densities with natural parameters $\mathbf{Y}$, $\hbox{Pois}(\mathbf{Z} ; \mathbf{Y}, \mathbf{b})$. By exponentiating the Poisson likelihood by a vector of weights, $\mathbf{W}$, the pseudo-likelihood contribution to the posterior is then proportional to
\begin{equation*}
    \prod_{i \in \mathcal{S}} \hbox{exp}\left\{W_i Z_i Y_i - W_i b_i  \hbox{exp}(Y_i)  \right\} = 
    \hbox{exp}\left\{(\bm{W} \odot \bm{Z})' \bm{Y} - (\bm{W} \odot \mathbf{b})' \hbox{exp}(\bm{Y})  \right\},
\end{equation*} with $\odot$ representing a Hadamard product, or element-wise multiplication. This is the same form as $\hbox{Pois}(\mathbf{Z}^*; \mathbf{Y}, \mathbf{b}^*)$, where $\mathbf{Z}^* = \mathbf{W} \odot \mathbf{Z}$ and  $\mathbf{b}^* = \mathbf{W} \odot \mathbf{b}$, and thus the MLG class of distributions is conjugate with pseudo-likelihoods built upon the Poisson distribution. This is important, as it allows us to use Gibbs sampling with conjugate full conditional distributions in order to sample from the posterior distributions.

Furthermore, \citet{brad18} show that the $\hbox{MLG}(\mathbf{c}, \alpha^{1/2}\mathbf{V}, \alpha \mathbf{1}, \alpha \mathbf{1})$ converges in distribution to a multivariate normal distribution with mean $\mathbf{c}$ and covariance matrix $\mathbf{V}$ as the value of $\alpha$ approaches infinity. This is convenient as it allows one to effectively use a latent Gaussian process model structure, while still maintaining the computationally benefits of conjugacy offered by the conjugate multivariate distribution theory. Herein, for illustration purposes, we use this type of prior distribution to approximate a latent Gaussian process. However, if desired, one could further model the shape and scale parameters from the MLG prior distribution, which can result in a more flexible shape to the posterior distribution.

We now consider the pseudo-likelihood Poisson multivariate log-Gamma model (PL-PMLG),
    \begin{equation}
    \begin{split}
        \mathbf{Z} | \bm{\eta, \beta, \xi} & \propto \prod_{\ell=1}^L \prod_{i \in S} \hbox{Pois}\left(Z_i^{(\ell)} | \lambda=Y_i^{(\ell)} \right)^{\stackrel{\sim}{w}_i} \\
        \hbox{log}(Y_i^{(\ell)}) &= \mathbf{x_i'}^{(\ell)} \bm{\beta} + \bm{\psi_i'} \bm{\eta}  + \xi_i^{(\ell)}, \; i \in \mathcal{S}, \; \ell=1,\ldots,L \\
        \bm{\eta}|\sigma_k & \sim \hbox{MLG}(\mathbf{0_r}, \alpha^{1/2}\sigma_k \mathbf{I_r}, \alpha \mathbf{1_r},\alpha \mathbf{1_r} ) \\
        \bm{\xi}^{(\ell)}|\sigma_{\xi} & \stackrel{ind.}{\sim} \hbox{MLG}(\mathbf{0_n}, \alpha^{1/2}\sigma_{\xi} \mathbf{I_n}, \alpha \mathbf{1_n},\alpha \mathbf{1_n} ), \; \ell = 1,\ldots,L \\
        \bm{\beta} & \sim  \hbox{MLG}(\mathbf{0_p}, \alpha^{1/2}\sigma_{\beta} \mathbf{I_p}, \alpha \mathbf{1_p},\alpha \mathbf{1_p} ) \\
        \frac{1}{\sigma_k} & \sim \hbox{Log-Gamma}^+(\omega, \rho) \\
        \frac{1}{\sigma_{\xi}} & \sim \hbox{Log-Gamma}^+(\omega, \rho), 
        \quad \sigma_{\beta}, \alpha, \omega, \rho >0,
    \end{split} 
\end{equation} where $Z_i^{(\ell)}$ is the $\ell$th response variable for unit $i$ in the sample. This model uses a pseudo-likelihood to account for informative sampling, and is built upon a Poisson response type in order to handle count valued survey data. In this work, the vector $\bm{\psi}_i$ corresponds to an incidence vector for which areal unit $i$ resides in. As such, the vector $\bm{\eta}$ acts as area level random effects, which are shared across response types in order to induce multivariate dependence. We note that this model is written for multivariate responses, but we focus only on a univariate example in this work. The parameters $\xi_i^{(\ell)}$ act as unit level random effects, and can account for fine scale variation due to missing unit level covariates. Finally, $\bm{\beta}$ corresponds to fixed effects, for which covariates may or may not be shared across response types. We place log-Gamma priors truncated below at zero (denoted $\hbox{Log-Gamma}^+$) on the parameters $1/\sigma_k$ and $1/\sigma_{\xi}$. This is done to maintain conjugate full conditional distributions, although other prior distributions could be used here with minimal tuning required as these are low-dimensional parameters deep in the model hierarchy. We set $\alpha=1000$ in order to approximate Gaussian prior distributions. We also set $\sigma_{\beta}, \omega, \rho = 1000$ in order to create vague prior distributions. However, if prior knowledge on these parameters exists, these values could be adjusted accordingly. The full conditional distributions used for Gibbs sampling can be found in Appendix~\ref{app1}.

\subsection{Boundary Correction}
One technical issue that arises when using a conjugate multivariate hierarchical modeling framework concerns data that are observed on the boundary of their support (i.e. zero counts for Poisson data). When zero counts are observed with Poisson data, the result is a full conditional distribution with a shape parameter of zero which is not well defined. Because the conjugate multivariate framework was only recently developed, there is relatively little literature on handling these boundary issues; however \citet{brad17} suggest using adjusted data, $Z_i^*=Z_i + c, \; i=1,\ldots,n$, by adding a small constant. This can work in many cases, depending on the dataset and the value of $c$, but is effectively sampling from an approximation to the posterior distribution.

Rather than sample from an approximate distribution, we use importance sampling to sample from the true posterior distribution, similar to the work of \citet{kim98}. In this case, the importance weights are proportional to the ratio of the adjusted pseudo-likelihood to the true pseudo-likelihood. However, with large sample sizes, the adjusted pseudo-likelihood can diverge from the true pseudo-likelihood. To this effect, we run a pilot chain (using 100 iterations) to find the average ratio of the true log-pseudo-likelihood to the adjusted log-pseudo-likelihood. We then scale the weights in the adjusted pseudo-likelihood by this average ratio. This has the effect of centering the adjusted pseudo-likelihood around the true pseudo-likelihood. The importance weights, taken at each iteration of the Gibbs sampler, are then proportional to $$
\prod_{i \in \mathcal{S}}
\frac{ \mbox{Pois}(Z_i | \cdot)^{\tilde{w}_i}}{ \mbox{Pois}(Z_i + c | \cdot)^{\tilde{w}_i^*}},
$$ where $\tilde{w}_i^*$ represents the scaled survey weight after multiplying by the average ratio mentioned above. We found that for the constant $\mathit{c},$ a value of one or two was ideal, as it minimized the extent of the divergence from the true pseudo-likelihood to the approximate one.

\section{Empirical Simulation Study}\label{sec:sim}
The American Community Survey (ACS) is an ongoing survey, with approximately 3.5 million households sampled annually, that is critical for informing how federal funds should be allocated. Although the complete microdata is not available to the public, public use microdata samples (PUMS) are available. PUMS only contain geographic indicators at the public use microdata area level (PUMA), which are aggregated areas such that each contains at least a population of 100,000 people. 
For this survey as well as others, vacant houses can pose a challenge when conducting the survey. \citet{brad17} use a simple random sample of ACS PUMS data within a single PUMA to predict housing vacancies by modeling the number of people per household as a Poisson random variable. This work illustrates the capacity of unit-level models to predict housing vacancies, however, because they used simple random sampling within a single PUMA, the methodology cannot be applied in an informative sampling context for SAE.

We construct an empirical simulation study to illustrate how the PL-PMLG can be used to create small area estimates of the number of housing vacancies. Using the state of Alabama, we treat the entire 2017 PUMS housing dataset as our population (or ``truth"). This dataset contains roughly 22,500 observations across 34 different PUMAs. We further subsample this data using the Midzuno probability proportional to size method \citep{midz51} within the `sampling' R package \citep{samp} , which we use to create our estimates. We then compare these estimates to the truth.

In addition to comparing the PL-PMLG to a direct estimator, we also wish to compare to another model based estimator. In this scenario, many of the direct estimators are equal to zero, which makes area-level modeling prohibitively difficult. Instead, because count data are often modeled as Gaussian on the log scale, we compare to a unit level model taking this approach. Because the data contains zero counts, a small constant, $\delta$, must be added to the data before taking the log transformation, and this transformation is undone when predictions are made. The full model hierarchy, which we call the Gaussian Approximation model (GA), is
        \begin{equation}
    \begin{split}
        \hbox{log}(Z_i + \delta) &\propto \hbox{N}(\bm{x_i' \beta} + \bm{\psi_i' \eta}, \sigma^2_{\xi})^{\stackrel{\sim}{w}_i}, \; i \in \mathcal{S} \\
        \bm{\eta} &\sim \hbox{N}(\bm{0}, \sigma^2_{\eta} \bm{I}) \\
        \bm{\beta} &\sim \hbox{N}(\bm{0}, \sigma^2_{\beta} \bm{I}) \\
        \sigma^2_{\xi} &\sim \hbox{IG}(\alpha_{\xi}, \kappa_{\xi}) \\
        \sigma^2_{\eta} &\sim \hbox{IG}(\alpha_{\eta}, \kappa_{\eta}) \\
        \sigma^2_{\beta}&, \alpha_{\eta}, \alpha_{\xi}, \kappa_{\eta}, \kappa_{\xi} > 0, 
    \end{split} 
\end{equation} where we use the vague prior distribution $\sigma^2_{\beta}=1000$, and $\alpha_{\eta}, \alpha_{\xi}, \kappa_{\eta}, \kappa_{\xi}=0.1$. We again use a pseudo-likelihood approach here in order to account for informative sampling. The rest of the model consists of fairly standard Bayesian mixed effects regression. We tested the value $\delta$ fixed over the values of (0.1, 1, 5), and found that $\delta=5$ yielded substantially lower MSE and bias for this example, which is what we present here.

For this simulation, we take a sample size of 5,000 from the PUMS data with probability proportional to $w_i$ (i.e., probability inversely proportional to the original probability of selection). We show that sampling this way induces informativeness by comparing to the unweighted version of our model. Our fixed effects consist of an intercept, and the number of bedrooms in the household, which we treat as a categorical variable. We calculate the  Horvitz-Thompson estimate (direct estimate) as well as the two model-based estimates. Finally, we repeat the process 50 times in order to compare MSE and absolute bias. For the PL-PMLG, unweighted PMLG (UW-PMLG) and GA estimators, we used Gibbs sampling for 2,000 iterations, discarding the first 1,000 as burn-in. Convergence was assessed visually through traceplots of the sample chains, and no lack of convergence was detected. We also compare to a Horvitz-Thompson direct estimator, with Hajek variance estimates using the \texttt{mase} package in \texttt{R} \citep{mase}.

\begin{figure}[H]
\centerline{\includegraphics[width=400pt,height=20pc]{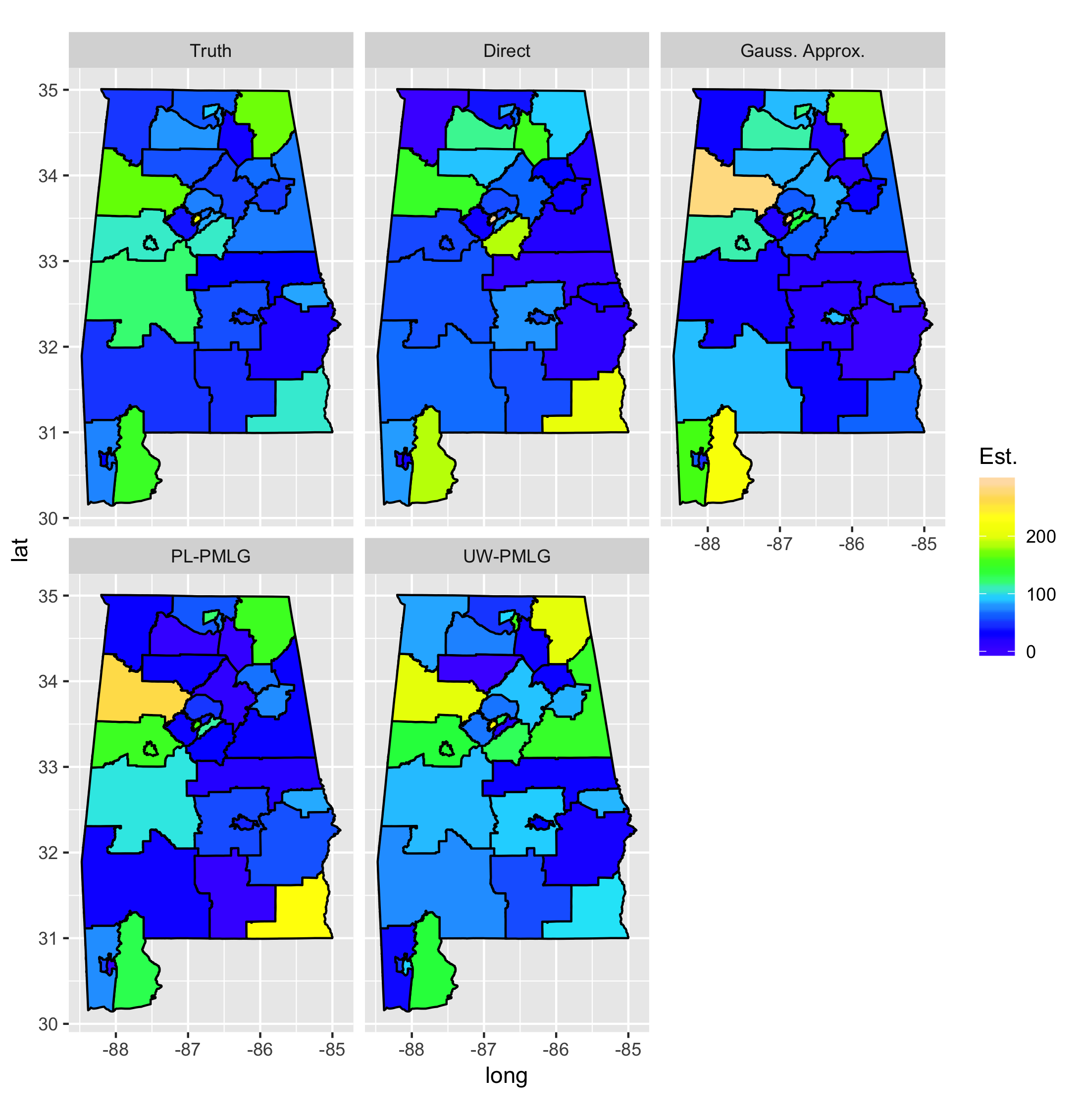}}
\caption{Point estimates of the number of housing vacancies by PUMA based on a single run of the simulation study.\label{point}}
\end{figure}

\begin{figure}[H]
\centerline{\includegraphics[width=400pt,height=20pc]{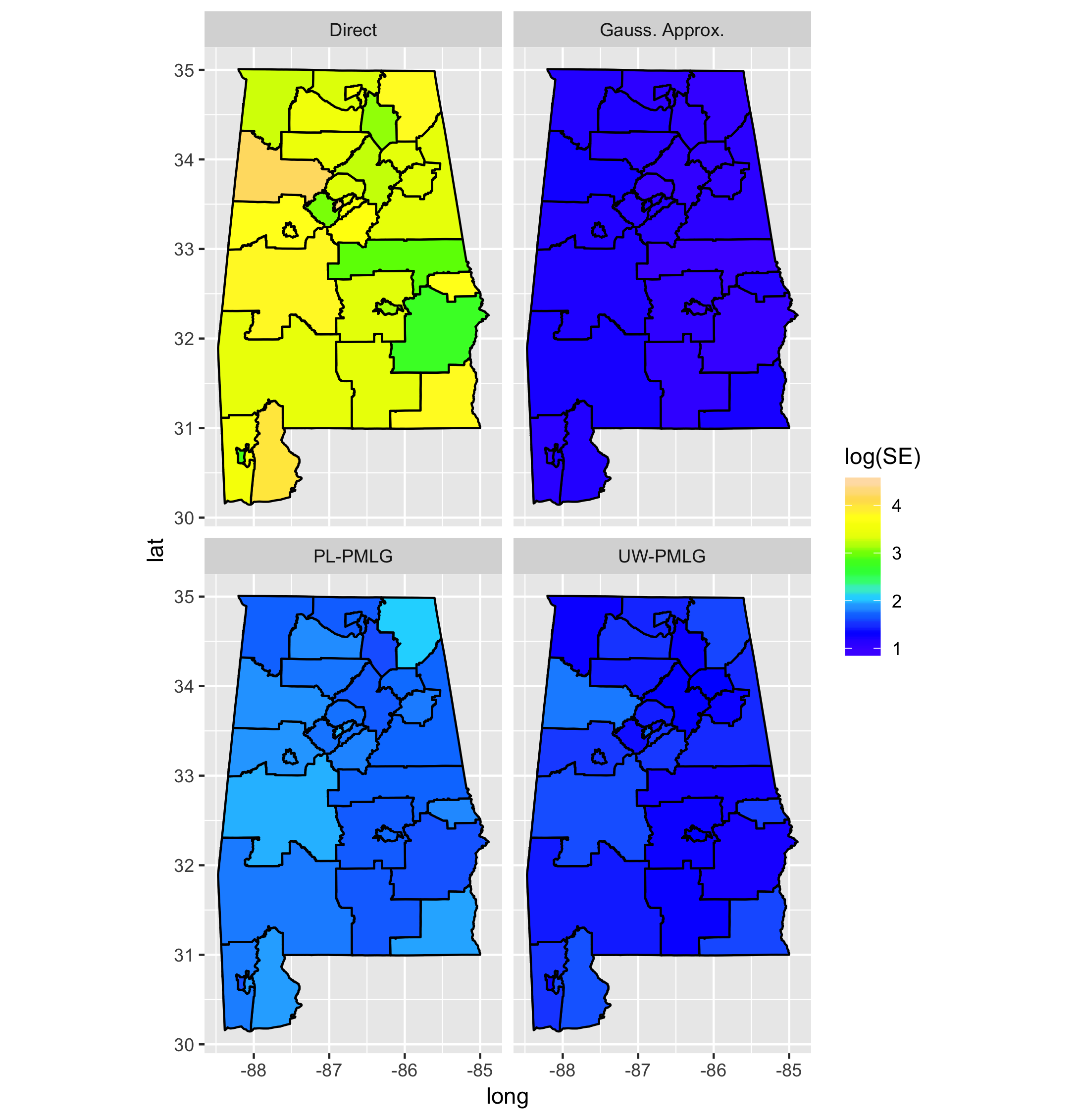}}
\caption{Standard error for the estimate of the number of housing vacancies by PUMA averaged over the simulation runs.\label{SE}}
\end{figure}

A summary of the simulation results can be found in Table~\ref{tab1}, where we compare the MSE and absolute bias of the PUMA level estimates for the total number of vacant housing units. The GA model does not provide a reduction in MSE compared to the direct estimator; however, the unweighted and PL-PMLG models do (12\% and 49\% respectively). Additionally, the absolute bias for the PL-PMLG is substantially lower than the GA and unweighted models. The significant reduction in MSE and bias comparing the PL-PMLG and UW-PMLG models indicates that there was an informative design, and the PL approach helps to account for this design. We also show the point estimates from a randomly chosen single run of the simulation under each estimator in Figure~\ref{point}. All of the estimators seem to capture the same general spatial trend, however the PL-PMLG estimator seems to most closely resemble the truth. As a final comparison, we plot the standard error of the estimates averaged across the 50 simulations on the log scale in Figure~\ref{SE}. To construct this figure, we compute a standard error of the estimate under each approach, for each of the 50 simulated datasets. For the model-based estimates, this standard error is the posterior predictive standard deviation. We then average these standard errors across the simulated data sets, in order to illustrate the expected uncertainty associated with each reported estimate. In some cases, the standard error of the direct estimate could not be obtained due to a point estimate of zero, in which case they have been removed from the average. As expected, the standard errors are dramatically lower for the model-based estimators than the direct estimator. In general the GA standard errors are slightly lower than the PL-PMLG, however the GA exhibits much higher MSE due to the increased bias, as evidenced by Table \ref{tab1}. Thus, the PL-PMLG appears to be a superior estimator overall.

\begin{center}
\begin{table}[t]%
\centering
\caption{Simulation results.\label{tab1}}%
\begin{tabular*}{500pt}{@{\extracolsep\fill}lcccc@{\extracolsep\fill}}
\hline
\textbf{Estimator} & \textbf{MSE}  & \textbf{Abs. Bias}   \\
\hline
Direct & 2250  & 3.5     \\
GA & 2526  & 33.9    \\
PL-PMLG & 1151  & 23.5     \\
UW-PMLG & 1983    & 32.7     \\
\hline
\end{tabular*}

\end{table}
\end{center}

\section{Discussion}\label{sec:disc}
There is a strong need for unit-level models that can handle survey data. Accounting for informative sampling design and modeling non-Gaussian data types are two of the biggest challenges in this setting. In this work, we present a new method for modeling count data while accounting for informative sampling. This method can be used for SAE as well as for more general modeling purposes. Our method relies on conjugate multivariate distribution theory, and we show that conjugacy is maintained when using a psuedo-likelihood approach to account for the survey design. We also extend the work of \citet{brad17} to handle the issue of zero counts through importance sampling.

Our approach is illustrated on a simulation study built upon public-use ACS data. This is a count data example where area-level models are not feasible and Gaussian models are not appropriate. Furthermore, this is an example where direct estimators are not useful due to excessively large MSE and standard errors. Our PL-PMLG approach is able to accurately estimate population quantities based on count variables while still maintaining computational efficiency.

There still remains further work to be done in the area of non-Gaussian survey data. Other data types such as binary random variables are prevalent and should be considered \citep{bau18, luery11}. The conjugate multivariate framework offered by \citet{brad17} has the potential to fit these types of data, although the boundary value issue may pose a computational challenges. Our importance sampling approach  works well in the Poisson case, but a more general solution may be attainable. Finally, non-Gaussian data should be explored in regards to other solutions to the informative sampling problem. The pseudo-likelihood approach may be one of the most popular approaches to informative sampling, but other methods exist and may yield additional gains in terms of precision.


\section*{Acknowledgments}

This report is released to inform interested parties of ongoing research and to encourage discussion of work in progress.  The views expressed are those of the authors, and not those of the U. S. Census Bureau.  This research was partially supported by the U.S. National Science Foundation (NSF) and the U.S. Census Bureau under NSF grant SES-1132031, funded through the NSF-Census Research Network (NCRN) program and NSF SES-1853096. Partial support of this work through the U.S. Census Bureau Dissertation Fellowship Program is gratefully acknowledged.

\pagebreak

\appendix
\section{Full Conditional Distributions for the PL-PMLG Model}\label{app1}

\subsection{Random Effects}
\begin{equation*}
    \begin{split}
        \bm{\eta} | \cdot &\propto \prod_{\ell=1}^L \prod_{i \in \mathcal{S}} \hbox{exp}\left\{\tilde{w}_i z_i^{(\ell)} \bm{\psi_i'}^{(\ell)} \bm{\eta} - \tilde{w}_i \hbox{exp}(\bm{x_i'}^{(\ell)} \bm{\beta} + \xi_i^{(\ell)})' \hbox{exp}(\bm{\psi_i'}^{(\ell)} \bm{\eta}) \right\} \\& \quad \times 
        \hbox{exp}\left\{\alpha \bm{1_r'} \alpha^{-1/2} \frac{1}{\sigma_k} \bm{I_r \eta} - \alpha \bm{1_r'} \hbox{exp}\left(\alpha^{-1/2} \frac{1}{\sigma_k} \bm{I_r \eta}\right)  \right\} \\
        &= \hbox{exp}\left\{\bm{\alpha_{\eta}' H_{\eta} \eta} - \bm{\kappa_{\eta}'} \hbox{exp}(\bm{H_{\eta} \eta})\right\} \\
        & \bm{H_{\eta}} = \left[
                            \begin{array}{c}
                            \bm{\Psi}  \\
                            \alpha^{-1/2} \frac{1}{\sigma_k} \bm{I_r} 
                            \end{array}
                            \right], \quad 
        \bm{\alpha_{\eta}} = (\bm{\tilde{w}'} \odot \bm{Z'}, \alpha \bm{1_r'})', \quad
        \bm{\kappa_{\eta}} = (\bm{\tilde{w}'} \odot \hbox{exp}(\bm{X \beta + \xi})', \alpha \bm{1_r'})' \\
        \bm{\eta} | \cdot &\sim \hbox{cMLG}(\bm{H_{\eta}}, \bm{\alpha_{\eta}}, \bm{\kappa_{\eta}})
    \end{split}
\end{equation*}

\begin{equation*}
    \begin{split}
        \bm{\xi} | \cdot &\propto \prod_{\ell=1}^L \prod_{i \in \mathcal{S}} \hbox{exp}\left\{\tilde{w}_i z_i^{(\ell)} \xi_i^{(\ell)} - \tilde{w}_i \hbox{exp}(\bm{x_i'}^{(\ell)} \bm{\beta} + \bm{\psi_i'}^{(\ell)} \bm{\eta})' \hbox{exp}(\xi_i^{(\ell)}) \right\} \\& \quad \times 
        \hbox{exp}\left\{\alpha \bm{1_n'} \alpha^{-1/2} \frac{1}{\sigma_{\xi}} \bm{I_n \xi} - \alpha \bm{1_n'} \hbox{exp}\left(\alpha^{-1/2} \frac{1}{\sigma_{\xi}} \bm{I_n \xi}\right)  \right\} \\
        &= \hbox{exp}\left\{\bm{\alpha_{\xi}' H_{\xi} \xi} - \bm{\kappa_{\xi}'} \hbox{exp}(\bm{H_{\xi} \xi})\right\} \\
        & \bm{H_{\xi}} = \left[
                            \begin{array}{c}
                            \bm{I_n}  \\
                            \alpha^{-1/2} \frac{1}{\sigma_{\xi}} \bm{I_n} 
                            \end{array}
                            \right], \quad 
        \bm{\alpha_{\xi}} = (\bm{\tilde{w}'} \odot \bm{Z'}, \alpha \bm{1_n'})', \quad
        \bm{\kappa_{\xi}} = (\bm{\tilde{w}'} \odot \hbox{exp}(\bm{X \beta + \Psi \eta})', \alpha \bm{1_n'})' \\
        \bm{\xi} | \cdot &\sim \hbox{cMLG}(\bm{H_{\xi}}, \bm{\alpha_{\xi}}, \bm{\kappa_{\xi}})
    \end{split}
\end{equation*}

\subsection{Fixed Effects}
\begin{equation*}
    \begin{split}
        \bm{\beta} | \cdot &\propto \prod_{\ell=1}^L \prod_{i \in \mathcal{S}} \hbox{exp}\left\{\tilde{w}_i z_i^{(\ell)} \bm{x_i'}^{(\ell)} \bm{\beta} - \tilde{w}_i \hbox{exp}(\bm{\psi_i'}^{(\ell)} \bm{\eta} + \xi_i^{(\ell)})' \hbox{exp}(\bm{x_i'}^{(\ell)} \bm{\beta}) \right\} \\& \quad \times 
        \hbox{exp}\left\{\alpha \bm{1_p'} \alpha^{-1/2} \frac{1}{\sigma_{\beta}} \bm{I_p \beta} - \alpha \bm{1_p'} \hbox{exp}\left(\alpha^{-1/2} \frac{1}{\sigma_{\beta}} \bm{I_p \beta}\right)  \right\} \\
        &= \hbox{exp}\left\{\bm{\alpha_{\beta}' H_{\beta} \beta} - \bm{\kappa_{\beta}'} \hbox{exp}(\bm{H_{\beta} \beta})\right\} \\
        & \bm{H_{\beta}} = \left[
                            \begin{array}{c}
                            \bm{X}  \\
                            \alpha^{-1/2} \frac{1}{\sigma_{\beta}} \bm{I_p} 
                            \end{array}
                            \right], \quad 
        \bm{\alpha_{\beta}} = (\bm{\tilde{w}'} \odot \bm{Z'}, \alpha \bm{1_p'})', \quad
        \bm{\kappa_{\beta}} = (\bm{\tilde{w}'} \odot \hbox{exp}(\bm{\Psi \eta + \xi})', \alpha \bm{1_p'})' \\
        \bm{\beta} | \cdot &\sim \hbox{cMLG}(\bm{H_{\beta}}, \bm{\alpha_{\beta}}, \bm{\kappa_{\beta}})
    \end{split}
\end{equation*}

\subsection{Variance Parameters}
\begin{equation*}
    \begin{split}
        \frac{1}{\sigma_k} | \cdot &\propto \hbox{exp}\left\{\alpha \bm{1_r'} \alpha^{-1/2} \frac{1}{\sigma_k} \bm{I_r \eta} - \alpha \bm{1_r'} \hbox{exp}\left(\alpha^{-1/2} \frac{1}{\sigma_k} \bm{I_r \eta}\right)  \right\} \\
        & \quad \times \hbox{exp}\left\{\omega \frac{1}{\sigma_k} - \rho \, \hbox{exp}\left(\frac{1}{\sigma_k}\right)\right\} \times I(\sigma_k > 0) \\
        & = \hbox{exp} \left\{\bm{\omega_k' H_k} \frac{1}{\sigma_k} - \bm{\rho_k'} \hbox{exp} \left( \bm{H_k} \frac{1}{\sigma_k} \right) \right\} \times I(\sigma_k > 0) \\
        & \bm{H_k} = (\alpha^{-1/2} \bm{\eta'}, 1)' \quad
        \bm{\omega_k} = (\alpha \bm{1_r'}, \omega)' \quad 
        \bm{\rho_k} = (\alpha \bm{1_r'}, \rho)'\\
        \frac{1}{\sigma_k} | \cdot & \sim \hbox{cMLG}(\bm{H_k, \omega_k, \rho_k}) \times I(\sigma_k > 0) 
    \end{split}
\end{equation*}

\begin{equation*}
    \begin{split}
        \frac{1}{\sigma_{\xi}} | \cdot &\propto \hbox{exp}\left\{\alpha \bm{1_n'} \alpha^{-1/2} \frac{1}{\sigma_{\xi}} \bm{I_n \xi} - \alpha \bm{1_n'} \hbox{exp}\left(\alpha^{-1/2} \frac{1}{\sigma_{\xi}} \bm{I_n \xi}\right)  \right\} \\
        & \quad \times \hbox{exp}\left\{\omega \frac{1}{\sigma_{\xi}} - \rho \, \hbox{exp}\left(\frac{1}{\sigma_{\xi}}\right)\right\} \times I(\sigma_{\xi} > 0) \\
        & = \hbox{exp} \left\{\bm{\omega_{\xi}' H_{\xi}} \frac{1}{\sigma_{\xi}} - \bm{\rho_{\xi}'} \hbox{exp} \left( \bm{H_{\xi}} \frac{1}{\sigma_{\xi}} \right) \right\} \times I(\sigma_{\xi} > 0) \\
        & \bm{H_{\xi}} = (\alpha^{-1/2} \bm{\xi'}, 1)' \quad
        \bm{\omega_{\xi}} = (\alpha \bm{1_n'}, \omega)' \quad 
        \bm{\rho_{\xi}} = (\alpha \bm{1_n'}, \rho)'\\
        \frac{1}{\sigma_{\xi}} | \cdot & \sim \hbox{cMLG}(\bm{H_{\xi}, \omega_{\xi}, \rho_{\xi}}) \times I(\sigma_{\xi} > 0) 
    \end{split}
\end{equation*}

\bibliographystyle{apacite}
\bibliography{main}%

\end{document}